\documentclass[aps,pra,twocolumn,superscriptaddress,showpacs,showkeys,amsmath,amssymb]{revtex4}

\usepackage{amsfonts}
\usepackage{amssymb,amsmath}
\usepackage{mathrsfs}
\usepackage{latexsym}
\usepackage{amsmath}
\usepackage[cp1251]{inputenc}
\usepackage{graphicx}
\usepackage{dcolumn}
\usepackage{bm}
\usepackage{color}

\RequirePackage{ifthen}
\RequirePackage[pdfstartview=FitH]{hyperref}
\begin{document}

    \title{Ground-state properties of dilute spinless fermions in fractional dimensions}
    
    \author{V.~Pastukhov\footnote{e-mail: volodyapastukhov@gmail.com}}
    \affiliation{Department for Theoretical Physics, Ivan Franko National University of Lviv, 12 Drahomanov Str., Lviv, Ukraine}

    \date{\today}


    \begin{abstract}
    We analyze zero-temperature universal properties of the simplest Galilean-invariant model of spinless low-dimensional fermions with short-range two-body interactions. In particular, it is shown that after proper renormalization of the coupling constant, even the dilute system possesses rich phase diagram that includes the superfluid state and the metastable `upper branch' behavior.
    \end{abstract}

    \maketitle

\section{Introduction}
\label{sec1}
\setcounter{equation}{0}
The idea of contact interaction is the cornerstone concept for understanding the quantum few-body physics. It is particularly useful in the theoretical analysis of the ultra-cold atomic gases, where diluteness of the system provides two substantial simplifications: i) the two-body interaction is the only important and ii) one can typically neglect effects of finiteness of the potential range. Both of them are compactly realized by the $\delta$-potential in low dimensions ($D<2$) and by the one-parameter $s$-wave pseudo-potential \cite{Huang} in $D<4$, which, however, can model the two-body interaction between bosons, or fermionic atoms in different spin (hyperfine) states. The problem of spinless fermions necessarily involves the momentum-dependent (pseudo-) interaction \cite{Derevianko,Idziaszek,Pricoupenko_06,Valiente_12} which in the most simple case can be restricted to the inclusion of $p$-wave channel. Recently, few- and many-body systems with the $p$-wave-type interaction between particles have been extensively studied in 1D (see \cite{Sowinski}, for review). However, the first attempts in this direction were stimulated by observation \cite{Cheon,Sen,Girardeau,Kanjilal,Brand} that the Lieb-Liniger model in the limit of strong inter-particle repulsion can be perturbatively described in terms of spinless fermions with $\delta''$-interaction. This `potential' is too singular even for 1D and should be properly treated by means of the tight-binding lattice \cite{Muth,Valiente_15,Valiente_18} or the effective field-theory \cite{Cui} regularization schemes. From practical point of view, the latter approach was shown to be extremely efficient for the derivation of the exact universal relations \cite{Sekino} and obtaining, by means of the two-channel model, first few terms in the high-momentum tail of the particle distribution \cite{Cui_Dong}. The characteristic feature of the $\delta''$-pseudo-potential in 1D is its dependence on a single parameter with dimension of length, which is clearly impossible for realization in all higher (integer) dimensions starting from 2D case \cite{Jiang}. The only exception is a `window' $1\le D<2$. The aim of the present study is to explore properties of spinless fermions with zero-range interactions in fractional dimensions. At present, this model is of a little practical use but seems to be interesting from the methodological point of view. Indeed, recalling an important aspect of the Bose-Fermi mapping in 1D -- the exact solvability \cite{Hao,Imambekov,Qi,Prem,Yin,Stouten}, where the only technical complication consists in the incorporation of the renormalization procedure in the Bethe ansatz (in other words how to construct the pseudo-potential and the appropriate two-body scattering matrix), we, therefore, have a system that because of the universal, far non-trivial phase diagram and the exact solution in 1D limit, is well-suited for testing various approximate approaches.

\section{Model}
We discuss the simplest model of interacting spinless (spin-polarized) fermions. More specifically, we consider a system with the following Euclidean action
\begin{eqnarray}\label{S}
S=\int dx \psi^*\{\partial_{\tau}-\xi\}\psi
-g\int dx (\nabla\psi^*)\psi^*\psi\nabla\psi,
\end{eqnarray} 
where $N$ particles with mass $m$ each, are assumed to be loaded in large volume $L^D$ (we are mostly interested in $1\le D<2$, but the extension to higher dimensions is straightforward) with periodic boundary conditions imposed. In (\ref{S}) the shorthand notations $x=(\tau,{\bf r})$, $\int dx \equiv \int^{\beta}_{0}d\tau \int_{L^D} d{\bf r}$, $\xi=-\frac{\hbar^2}{2m}\nabla^2-\mu$ are used, where $\beta$ and $\mu$ denote the inverse temperature and the chemical potential of Fermi gas fixing its density $n=N/L^D$, respectively. The minimal two-body local interaction preserving the Galilean invariance of the system is characterized by a single (bare) coupling constant $g$. This type of interaction is badly-defined in any spacial dimension $D$, therefore by writing it down in action $S$ we mean the situation, when fermions interact via some `physical' two-body potential, but its range $1/\Lambda$ is the smallest parameter with dimension of length. It is then believed that such an effective theory (with $g$ explicitly dependent on $\Lambda$) properly describes properties of the system at energy scales much smaller than $\hbar^2\Lambda^2/m$.

\begin{figure}[h!]
	\includegraphics[width=0.46\textwidth,clip,angle=-0]{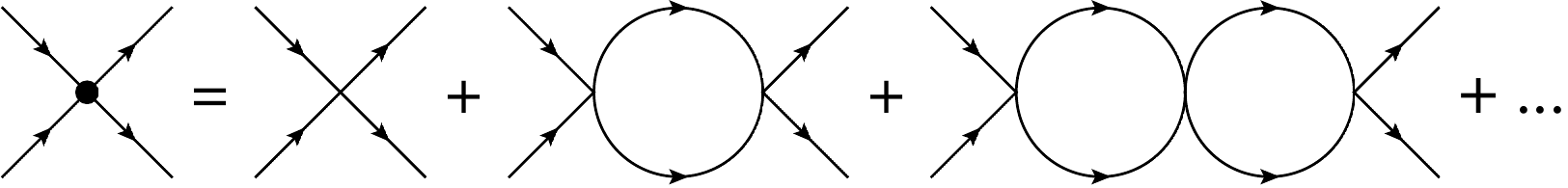}
	\caption{Diagrams contributing to vertex $\mathcal{T}_0(P_1,P_2|P'_2,P'_1)$.}
	\label{diagrams}
\end{figure}

In order to figure out the coupling-constant renormalization, let us consider the two-body problem with our potential (actually pseudo-potential). This can be easily done by considering the two-particle vertex function $\mathcal{T}(P_1,P_2|P'_2,P'_1)$ [from now on capital letters denote $D+1$-momenta $P=(\nu_p,{\bf p})$ of particles] in the zero-density ($\mu\to 0$) limit. Summation of all particle-particle diagrams (see Fig.~\ref{diagrams}) 
leads to the integral equation for $\mathcal{T}_0(P_1,P_2|P'_2,P'_1)=\delta_{P_1+P_2,P'_1+P'_2}{\bf p}_{12}{\bf p}'_{12}t_0(P_1+P_2)$ (here ${\bf p}_{12}={\bf p}_{1}-{\bf p}_{2}$), which reduces to the algebraic one for $t_0(P_1+P_2)$, with the solution
\begin{eqnarray}\label{t_0}
t^{-1}_0(Q)=g^{-1}
+\frac{1}{L^D}\sum_{{\bf p}}\frac{2p^2/D}{2\varepsilon_p+\varepsilon_{q}/2-i\omega_q},
\end{eqnarray} 
here $\varepsilon_p=\hbar^2p^2/2m$ is the one-particle dispersion. The above integral is infinite of course, but this divergence can be cured by the dimensional regularization procedure or by introducing the cutoff $\Lambda$ for momentum in addition to the `physical' coupling constant $g^{-1}_F=g^{-1}+\frac{1}{L^D}\sum_{{\bf p}}\frac{p^2}{D\varepsilon_p}$. In terms of renormalized coupling $g_F$, the function $t_0(P_1+P_2)$ contains all information about the two-body scattering in vacuum. Particularly, for all positive $g_F$s there is always one bound state $\varepsilon_b=-\hbar^2/(ma^2)$ of size $a$, which can be related to the magnitude of coupling constant
\begin{eqnarray}\label{g_F}
g_F=\frac{(4\pi)^{D/2}D/2}{\Gamma(1-D/2)}\frac{\hbar^2a^D}{m}.
\end{eqnarray} 
For calculations of the scattering amplitude (actually, the $p$-wave contribution, which is the only non-zero in our case), we have to perform the analytical continuation of $\mathcal{T}_0(P_1,P_2|P'_2,P'_1)$ in the upper complex half-plane and take its on-shell expression
\begin{eqnarray}\label{f_F}
f_F\propto \frac{k^2}{g^{-1}_F-\frac{\Gamma(1-D/2)}{(4\pi)^{D/2}D/2}\frac{mk^D}{\hbar^2}e^{-i\pi D/2}},
\end{eqnarray}
where $k$ is the modulus of transferred momentum. The connection to the system of point-like bosons, i.e., the Lieb-Liniger model can be naively tracked by considering the scattering of two Bose particles interacting via potential $g_B\delta({\bf r})$. Below $D=2$, this potential is well-defined, and a very similar, to the one described above, calculation procedure leads to the result (recall, this is the $s$-wave channel)
\begin{eqnarray}\label{f_B}
f_B\propto \frac{1}{g^{-1}_B-\frac{\Gamma(1-D/2)}{(4\pi)^{D/2}}\frac{mk^{D-2}}{\hbar^2}e^{-i\pi D/2}}.
\end{eqnarray}
Comparing $f_F$ and $f_B$ in 1D, we see that the denominators of these function are equal to each other if $g^{1D}_F=-2\hbar^4/(m^2g^{1D}_B)$ \cite{Cheon,Sen,Girardeau,Kanjilal,Brand}. Therefore, at \textit{all} couplings, the scattering properties of the 1D model (\ref{S}) are identical to those for a system of bosons with $\delta$-repulsion. Note that such a correspondence is intrinsic only for 1D case and does not hold for higher spacial dimensions at any finite $g_{B,F}$. When $g^{-1}_{B,F}\to 0$, however, the functional forms of two expressions are the same, hinting the equivalence of the Bose gas with infinite point-like repulsion between particles and the system of spinless fermions with short-ranged $p$-wave interaction at unitary. Although this conclusion has been drawn here for all dimensions $D<2$, the general tendency may potentially realize in higher $D$s.

\section{Dilute limit}

\begin{figure}[h!]
	\includegraphics[width=0.5\textwidth,clip,angle=-0]{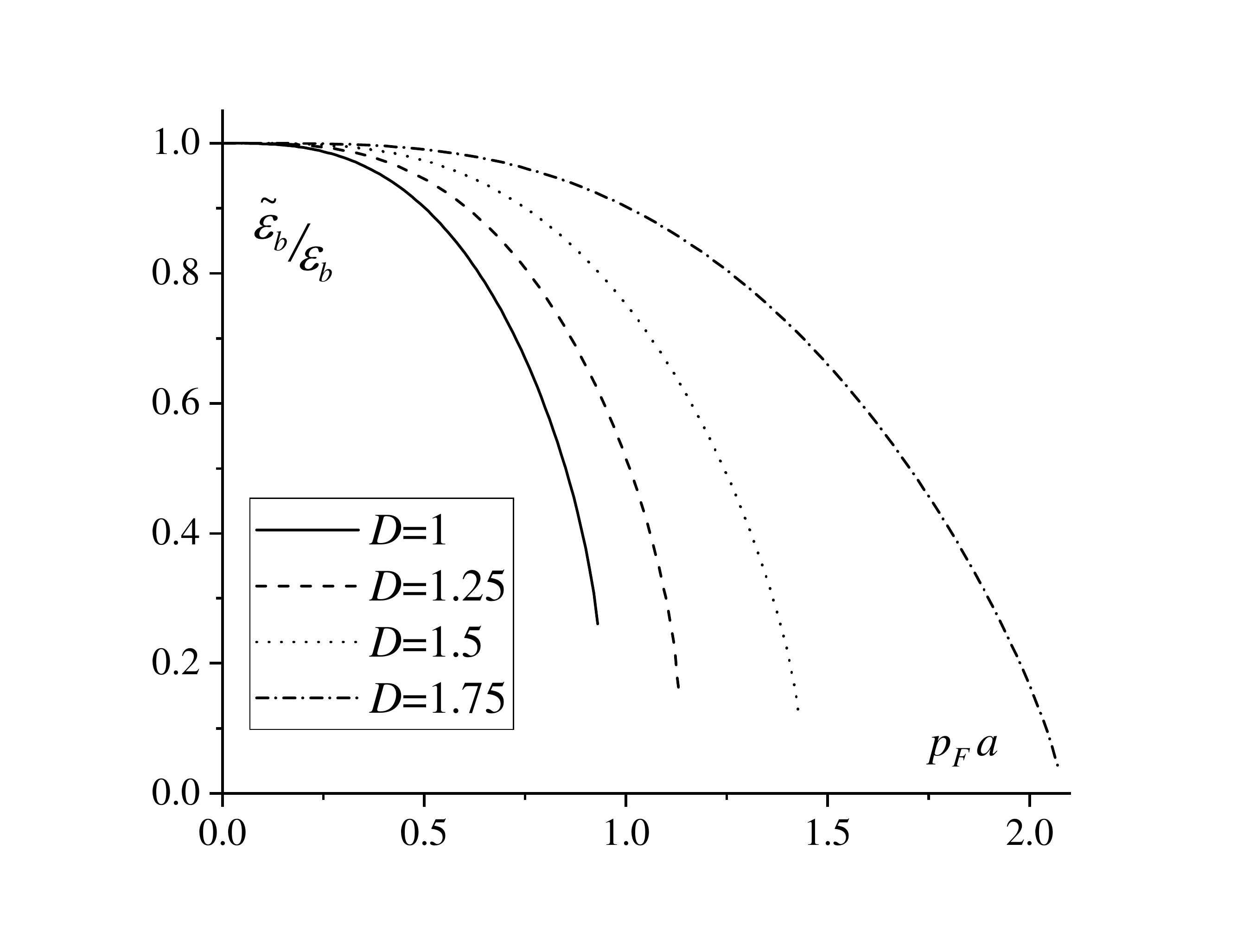}
	\caption{Two-body bound state energy (shifted on $2\mu$ and plotted in units of $\varepsilon_b$) at finite densities and zero center-of-mass momentum.}
	\label{bound_state_en}
\end{figure}

The effects of low densities can be easily captured by considering the two-body scattering processes (see Fig.~\ref{diagrams}) in the presence of the Fermi surface. The latter, in practice, means that the fermionic propagators are supplemented by non-zero chemical potential. In this approximation, the equation for vertex $\mathcal{T}(P_1,P_2|P'_2,P'_1)$ is more tricky nonetheless still tractable. The solution 
\begin{eqnarray}\label{Tau}
\mathcal{T}(P_1,P_2|S_2,S_1)=\delta_{P_1+P_2,S_1+S_2}p^i_{12}s^j_{12}t_{ij}(P_1+P_2),
\end{eqnarray}
(where sums over repeating indices $1\le i,j\le D$ are understood) is determined by the symmetric matrix of size $D\times D$ with the following elements:
\begin{eqnarray}\label{t_ij}
t^{-1}_{ij}(Q)=\frac{\delta_{ij}}{g}+\frac{2}{L^D}\sum_{{\bf p}}p^ip^j\frac{1-n_{{\bf p}+{\bf q}/2}-n_{{\bf p}-{\bf q}/2}}{\xi_{{\bf p}+{\bf q}/2}+\xi_{{\bf p}-{\bf q}/2}-i\omega_q},
\end{eqnarray}
where $n_{\bf p}=\Theta(p_F-p)$ is the Fermi distribution at absolute zero. We now see that because of the quantum-statistical effects, the relative motion of two particles is dependent on their center-of-mass momentum. But when total momentum of colliding particles is zero, matrix $t_{ij}(Q)_{{\bf q}=0}=\delta_{ij}t(i\omega_q)$ is isotropic. It is instructive to explore the role of finite densities in the fate of vacuum two-body states at positive $g_F$s. Having calculated real poles $\tilde{\varepsilon}_b$ of $t(\omega+i0)$ (see Fig.~\ref{bound_state_en}),
we have found out that the bound-state energy is strongly exhausted at finite $n$, and the region of existence of the two-body bound states for positive $g_F$s is always restricted by some maximal value of the dimensionless parameter $p_Fa$. At all negative couplings (it is convenient to parametrize $|g_F|$ for $g_F<0$ by length-scale $a$), contrarily, the Cooper phenomenon is observed with characteristic dependence of the energy gap $\Delta_{\textrm{gap}}/2\mu\propto \exp\left\{-\frac{\pi(p_Fa)^{-D}}{2\sin(\pi D/2)}\right\}$ on the coupling parameter in the limit $p_Fa\to 0$. 

Before we proceed to the calculation of many-body ground state energy, let us briefly summarize the phase diagram of the considered system. The case of positive $g_F$s, which will be referred below as a `repulsive' one, is characterized by the presence of two-body vacuum bound states, which, however, can disappear at finite densities of the system. But the most important thing about `repulsive' interaction is that it slightly increases the energy of two scattering particles. At negative couplings $g_F$, there are no vacuum bound states, but at any finite densities the system is expected to be in the superfluid phase. The latter fact is in agreement with the above-mentioned correspondence between our 1D model and the system of point-like bosons.

\subsection{Normal phase}
\begin{figure}[h!]
	\includegraphics[width=0.5\textwidth,clip,angle=-0]{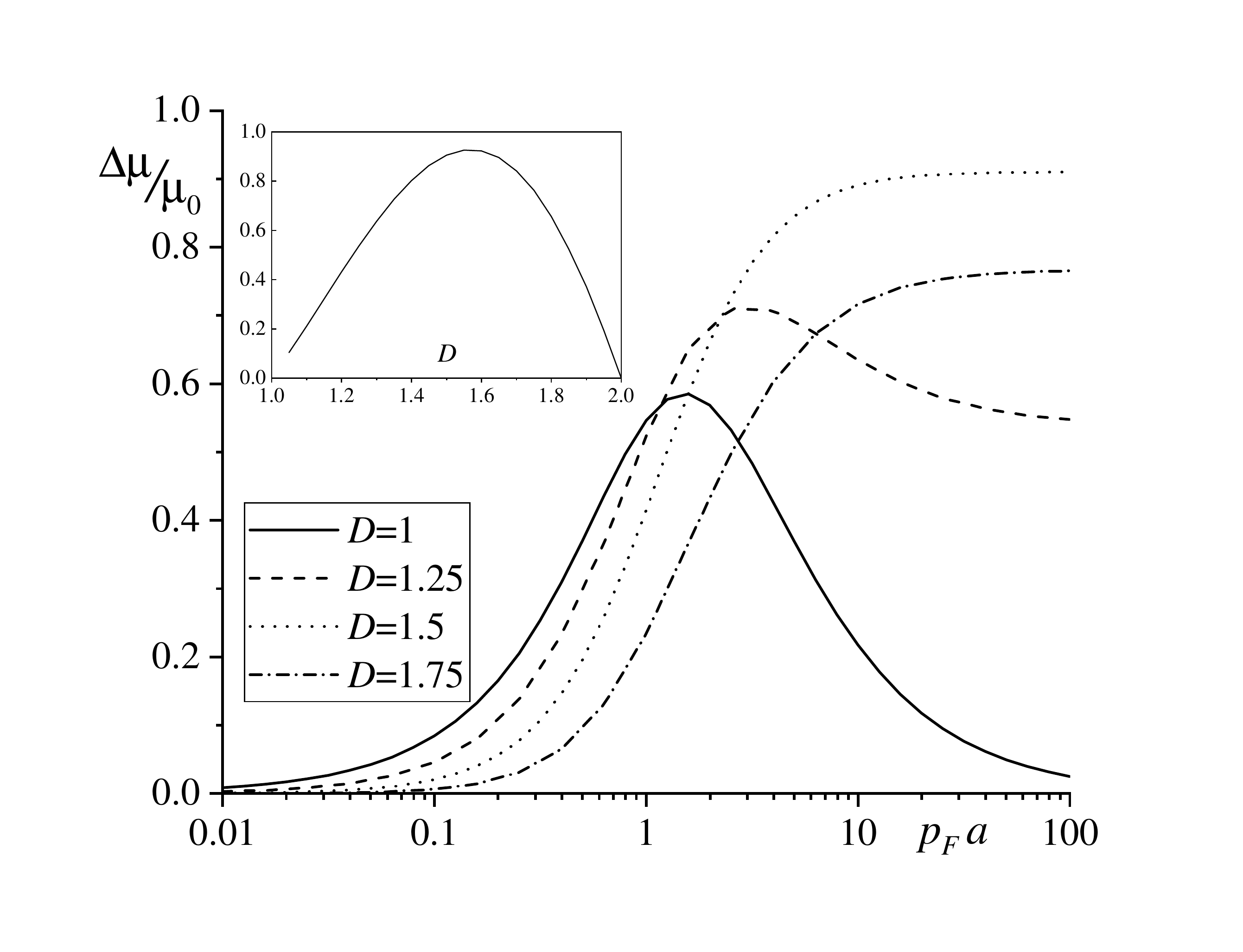}
	\caption{Correction to chemical potential of the `upper branch' repulsive Fermi gas (in units of $\mu_0$). Insert shows $\Delta\mu/\mu_0$ at unitary ($p_Fa\to \infty$) versus spacial dimension $D$.}
	\label{mu_rep}
\end{figure}

The simplest way to obtain energetics, in this case, is to use the Fermi liquid analogy. Indeed, in the renormalization-group sense, the $p$-wave interaction is marginally relevant even in $D<2$, therefore, the jump in the particle distribution survives for all coupling strengths and its position is the same for interacting and non-interacting systems. This observation allows to obtain chemical potential $\mu$ through the real part (after analytical continuation in upper complex half-plane) of the self-energy on the Fermi surface. The simplest correction to fermionic Green's function can be calculated by closing the loop around vertex (\ref{Tau})
\begin{eqnarray}\label{Sigma}
\Sigma(P)=\frac{1}{\beta L^D}\sum_{S}\mathcal{T}(P,S|S,P)\frac{1}{i\nu_s-\xi_s}.
\end{eqnarray}
Of course, the proper calculations with full vertex function (\ref{Tau}) and (\ref{t_ij}) are complicated and, therefore, left for future studies. Here we only restrict ourselves to the most simple approximation, namely, to the replacement $\mathcal{T}(P,S|S,P)\to \mathcal{T}_0(P,S|S,P)$ and neglecting the chemical potential in the denominator of (\ref{Sigma}) after frequency integration. The resulting formula
\begin{eqnarray}\label{Sigma_0}
\Sigma(P)|_{i\nu_p\to \xi_p+i0}=\frac{1}{L^D}\sum_{{\bf s}}\frac{g_Fs^2n_{{\bf p}-{\bf s}}}{1-\frac{g_F}{|g_F|}\left(\frac{sa}{2}\right)^{D}e^{-i\pi D/2}}\nonumber\\
-\Theta(g_F)\frac{1}{L^D}\sum_{{\bf s}}\frac{2}{D}\frac{g_Fs^2\Theta(|\varepsilon_b|-2\varepsilon_{{\bf p}-{\bf s}/2})}{1+\left(\frac{sa}{2}\right)^{2}},
\end{eqnarray} 
despite its simplicity, qualitatively correctly describes the properties of the system and has a clear structure: the first term comes from scattering states, while the second one is density-independent and originates from the two-body bound states. Being divided on energy $\varepsilon_b$ it can be treated as a fraction of atoms involved in the bound-state formation. Setting $p=p_F$ in Eq.~(\ref{Sigma_0}) and taking the real part of it, we obtain (see Fig.~\ref{mu_rep})
correction $\Delta \mu$ to the chemical potential of interacting fermions. It should be noted, that likewise 3D two-component fermions with short-ranged repulsion \cite{Chang,Shenoy}, only the so-called `upper branch' is of physical relevance. Indeed, lowering interaction strength $a\to 0$, we observe that the second term of energy enormously decreases (because $\varepsilon_b\to -\infty$ in this limit). But if fermions are initially prepared as a non-interacting gas and then the weak `repulsion' is suddenly switched on, the life-time of this metastable state is supposed to be very large before the system will `fall down' in its true ground state. Figure~\ref{mu_rep} reveals the non-monotonic dependence of the chemical-potential correction on $p_Fa$ in dimensions close to $D=1$. In the unitary limit, the behavior of $\Delta \mu$ as a function of $D$ is also quite unexpected, particularly, close to 2D the system of unitary spinless fermions is unaffected by $p$-wave interaction. The latter observation suggests the perturbative approach to the problem of unitary spinless Fermi gas (and, following our conjecture based on the comparison of scattering amplitudes $f_F$ and $f_B$, also unitary Bose gas with point-like repulsion) in terms of small parameter $\epsilon=2-D$.

At this point it is also interesting to consider the 1D limit of formula (\ref{Sigma}). Particularly, it contains correct information about properties of the system up to second order in the formal series expansion in powers of $g_F$
\begin{eqnarray}\label{Sigma_series}
\Sigma(P)=\frac{1}{L^D}\sum_{{\bf s}}g_F({\bf p}-{\bf s})^2n_{{\bf s}}-\frac{2}{L^{2D}}\sum_{{\bf s},{\bf q}}g^2_F({\bf p}-{\bf s},{\bf q})^2\nonumber\\
\times \left\{\frac{n_{{\bf s}}\left(1-n_{{\bf q}-\frac{{\bf p}+{\bf s}}{2}}-n_{{\bf q}+\frac{{\bf p}+{\bf s}}{2}}\right)+n_{{\bf q}-\frac{{\bf p}+{\bf s}}{2}}n_{{\bf q}+\frac{{\bf p}+{\bf s}}{2}}}{\xi_{{\bf q}-\frac{{\bf p}+{\bf s}}{2}}+\xi_{{\bf q}+\frac{{\bf p}+{\bf s}}{2}}-\xi_{{\bf s}}-i\nu_p}\right.\nonumber\\
\left.-\textrm{P.V.}\frac{n_{{\bf s}}}{2\varepsilon_{q}-\varepsilon_{{\bf p}-{\bf s}}/2}\right\}+\ldots.
\end{eqnarray}
Being calculated on the mass-shell $i\nu_p\to \xi_p+i0$ in 1D, the real part of the above expression was found to be amazingly simple
\begin{eqnarray}\label{Sigma_1D}
\Re\Sigma(P)|_{i\nu_p\to \xi_p+i0}=\frac{g_Fp^3_F}{\pi}\left\{\frac{1}{3}+\left(\frac{p}{p_F}\right)^2\right\}\nonumber\\
+\frac{3mg^2_Fp^4_F}{2\pi^2\hbar^2}\left\{\frac{2}{3}+\left(\frac{p}{p_F}\right)^2\right\}+\mathcal{O}(g^3_F),
\end{eqnarray}
and making the formal replacement $g^{1D}_F\to-2\hbar^4/(m^2g^{1D}_B)$ and using the Fermi-liquid prescription after, we obtain the chemical potential $\mu/\mu_0=1-\frac{16}{3\gamma}+\frac{20}{\gamma^2}+\ldots$ \cite{Zvonarev} and the quasiparticle effective mass $m/m^*=1-\frac{4}{\gamma}+\frac{12}{\gamma^2}+\ldots$ \cite{Ristivojevic} for the Lieb-Liniger model in the large-$\gamma$ (where $\gamma=mg_B/(\hbar^2n)$ is the dimensionless coupling parameter) limit. Therefore, even by using the incorrect (ideal-gas-like) Green's function, we can calculate the series expansion for thermodynamics of the point-like bosons. In fact, at negative $g_F$s the system possesses the superfluid behavior with the characteristic energy gap in the excitation spectrum. At small $g_F$ (large $\gamma$), however, the effects of the gap on thermodynamics of the system are exponentially suppressed.

\subsection{Superfluid phase}
\begin{figure}[h!]
	\includegraphics[width=0.5\textwidth,clip,angle=-0]{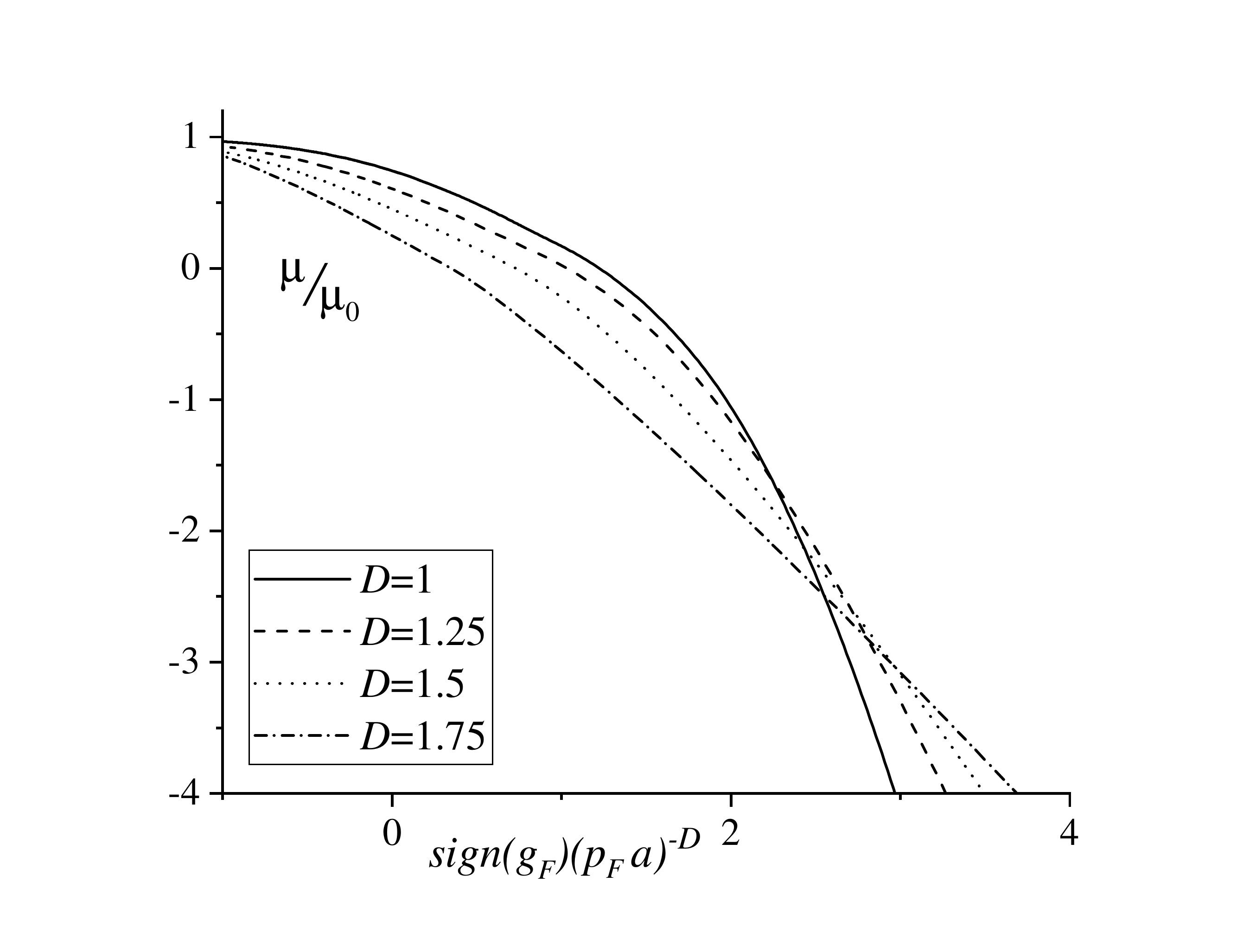}
	\caption{Dimensionless chemical potential $\mu/\mu_0$ of the superfluid fermions close to unitary. Note that in the crossover region chemical potential $\mu$ changes its sign.}
	\label{mu_superfluid}
\end{figure}

\begin{figure}[h!]
	\includegraphics[width=0.5\textwidth,clip,angle=-0]{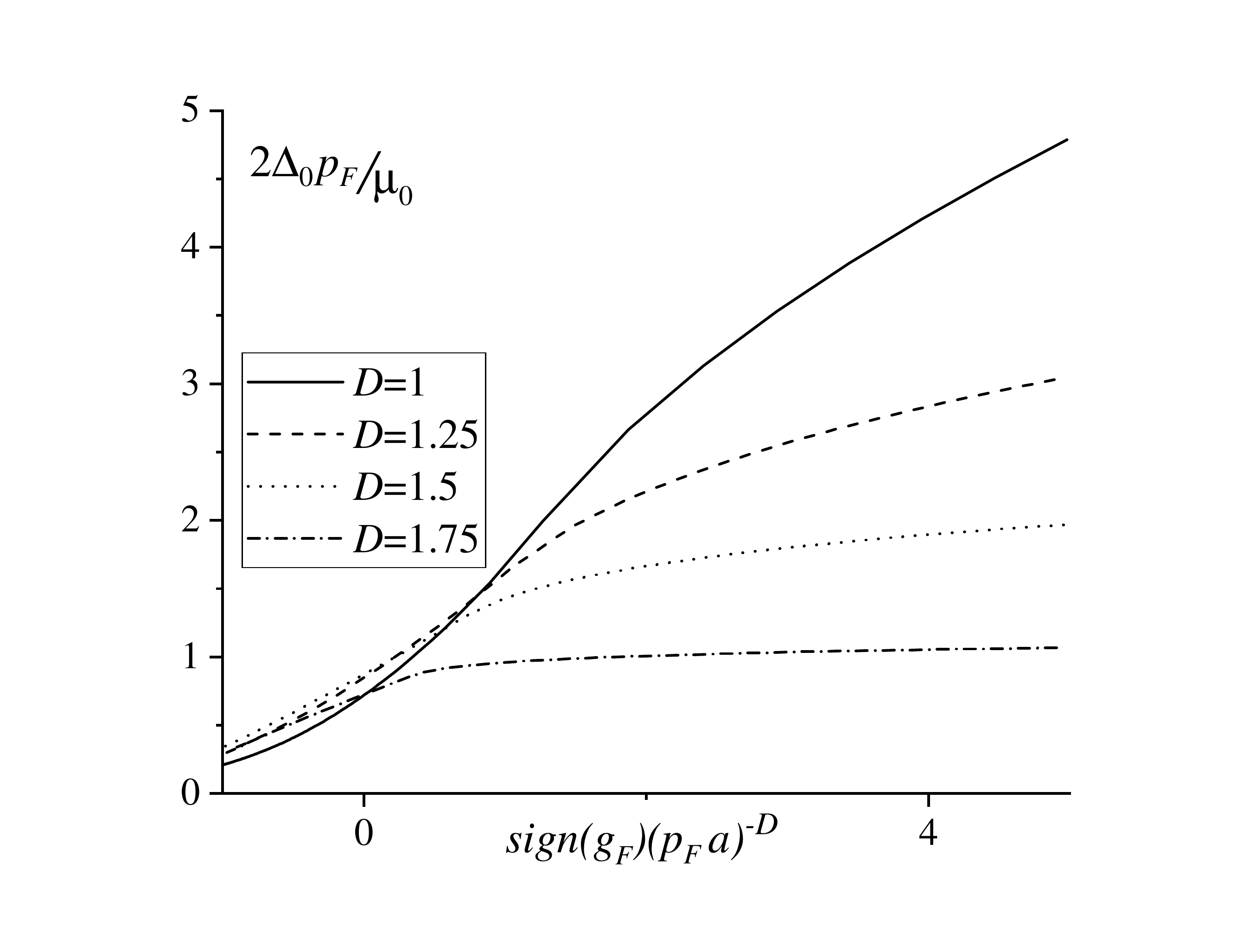}
	\caption{Dependence of the order parameter $\Delta_0$, which determines the energy gap (actually pseudo-gap in $D>1$) from the BCS side.}
	\label{Delta}
	\end{figure}

It is more convenient to discuss the attractive interaction with the transformed action instead of (\ref{S}). Introducing auxiliary complex $D$-component vector field ${\bf \Delta}^*(x)$, ${\bf \Delta}(x)$, one equivalently rewrites $S$ in the following way (it is assumed that $g<0$)
\begin{eqnarray}\label{S_Delta}
S=\int dx \psi^*\{\partial_{\tau}-\xi\}\psi
+g^{-1}\int dx {\bf \Delta}^*{\bf \Delta}\nonumber\\
-\int d x\left\{i{\bf \Delta}^*\psi\nabla\psi+\textrm{c.c.}\right\}.
\end{eqnarray} 
Then, by utilizing the standard prescription \cite{Gurarie} in our case, we can separate the non-zero expectation value of the field ${\bf \Delta}(x)={\bf \Delta}_0+\delta{\bf \Delta}(x)$ (note that $\int d{\bf r}\delta{\bf \Delta}(x)=0$, and only one component of the constant vector ${\bf \Delta}_0$ is not equal to zero identically). The physical meaning of the order parameter is different for $D=1$ and $D\neq 1$ cases. In 1D, $\Delta_0$ is the gap in the one-particle spectrum from the BCS side ($\mu<0$) of crossover, while in all higher dimensions it is a parameter that governs the degree of anisotropy of elementary excitations. Minimizing the grand potential with respect to ${\bf \Delta}_0$, we obtain the gap equation
\begin{eqnarray}\label{gap_Eq}
g^{-1}{\bf \Delta}_0
-\frac{i}{\beta L^D}\int dx \langle\psi\nabla\psi\rangle=0,
\end{eqnarray}
(here $\langle\ldots\rangle$ denotes the statistical averaging with action $S$) that should be supplemented with the equation for the average density of fermions
\begin{eqnarray}\label{n_Eq}
n=\frac{1}{\beta L^D}\int dx \langle\psi^*\psi\rangle.
\end{eqnarray}
In the following, we will use the mean-field approximation when one does not take into account fluctuation fields $\delta{\bf \Delta}^*(x)$, $\delta{\bf \Delta}(x)$. For the system under consideration, this approximation gives only a qualitative picture of its behavior because it totally neglects the normal self-energy insertion. Fortunately, the simplest inclusion of these terms in the fermionic propagator only shifts the chemical potential and renormalizes quasiparticle mass [note that the structure of the first term in Eq.~(\ref{Sigma_1D}) is universal, i.e., independent of the spacial dimension $D$] providing the qualitative correctness of the mean-field description. The straightforward calculations of averages yield
\begin{eqnarray}\label{psipsi}
	\langle\psi^*_P\psi_P\rangle=
	\frac{-i\nu_p-\xi_p}{\nu^2_p+E^2_p},\ \ 
	\langle\psi_{-P}\psi_{P}\rangle=\frac{2{\bf \Delta}_0{\bf p}}{\nu^2_p+E^2_p},
\end{eqnarray}
where $E^2_p=\xi^2_p+4|{\bf \Delta}_0{\bf p}|^2$.  After substitution in the system of coupled Eqs.~(\ref{gap_Eq}), (\ref{n_Eq}), the Matsubara frequency integration and renormalization of the bare coupling constant $g$, one obtains
\begin{eqnarray}
n=\frac{1}{2L^D}\sum_{{\bf p}}\left\{1-\frac{\xi_p}{E_p}\right\},
\end{eqnarray}
\begin{eqnarray}
g^{-1}_F\Delta^2_0+\frac{1}{L^D}\sum_{{\bf p}}\left\{\frac{({\bf \Delta}_0{\bf p})^2}{E_p}-\frac{\Delta^2_0p^2}{D\varepsilon_p}\right\}=0,
\end{eqnarray}
where ${\bf \Delta}_0$ is assumed to be real-valued. Besides the trivial solution ${\bf \Delta}_0=0$, these equations have the non-trivial one, which manifests the superfluid phase. Furthermore, they also demonstrate the type of BCS-BEC crossover, when $g^{-1}_F$ tends to zero from opposite sides. The results of the numerical solution for chemical potential in a few spacial dimensions are plotted in Figs.~\ref{mu_superfluid},\ref{Delta} .
For convenience, the coupling constant is parameterized, even in this `attractive' case, by a positive parameter $a$ as follows: $g_F=\textrm{sign}(g_F)\frac{(4\pi)^{D/2}D/2}{\Gamma(1-D/2)}\frac{\hbar^2a^D}{m}$. The physics behind the behavior of $\mu$ is very clear. At weak negative couplings, the gap is small, and the chemical potential is exponentially close to its ideal-gas value. Increase of $p_Fa$ (when $g_F<0$) changes this behavior, particularly, providing that  $\mu$ and $\Delta_0p_F$ are the same order magnitude at unitary. For small and positive $g_F$s the BCS-type pairing is replaced by the formation of molecules with large binding energies (this is actually the `lower branch' that was not discussed in the previous subsection). There is also a region $0<(p_Fa)^{-1}<2$, where the bound-states formation is exhausted by the many-body effects (see Fig.~\ref{bound_state_en}) and may be incorrectly treated by the mean-field approximation adopted here. In order to get more insight into the properties of the system in this region, one requires to include the quantum fluctuations. However, the impact of the beyond-mean-field effects on the superfluid phase of the considered model is out of scope of the present study, and together with the proper incorporation of the Gaussian fluctuations in the `upper branch'-like behavior, will be reported elsewhere.

\section{Concluding remarks}
In summary, we have studied the zero-temperature behavior of spinless fermions with minimal local two-body interactions between particles in the limit of extreme diluteness. It is shown that below two spacial dimensions properties of the system at fixed density are universal, i.e., controlled by a single parameter $a$, which determines width of the two-body vacuum bound state and plays a role of the $s$-wave scattering length simultaneously. Our results reveal a deep qualitative similarity between spinless fermions in dimensions below $D<2$ and spin-$1/2$ particles with short-ranged two-body interaction in $D<4$. Particularly, depending on a sign of the renormalized coupling constant $g_F$, the phase diagram of both systems includes the metastable `upper branch'-like behavior and the superfluid state with two different pairing mechanisms, namely, the Cooper instability and a formation of the two-atom molecules.  We also give some arguments that at unitary (from the BCS side) our model is equivalent to the system of bosons with infinite point-like repulsive two-body potential. This correspondence, which is well-known in 1D, can potentially allow the exploration of properties of low-dimensional $D<2$ strongly-interacting bosons in fermionic language. It is also interesting to explore the few-body physics of the considered model. This work is currently in progress, but even without knowing the exact solution it is understood that the few-particle spectrum should contain the cluster bound states at `repulsive' couplings.

\begin{center}
	{\bf Acknowledgements}
\end{center}	
We thank Prof.~Andrij~Rovenchak, Dr.~Orest~Hryhorchak for fruitful discussions and Dr.~Iryna~Pastukhova for careful reading of the manuscript. We are also grateful to Prof.~Mikhail~Zvonarev for sending copy of his Ph.D. Thesis.

\end{document}